\documentclass[aps,prl,preprint,tightenlines,superscriptaddress,showpacs,byrevtex]{revtex4}
\usepackage{graphicx}
\usepackage{dcolumn}
\usepackage{bm}

\begin{document}

\vspace*{-3\baselineskip}
\resizebox{!}{3cm}{\includegraphics{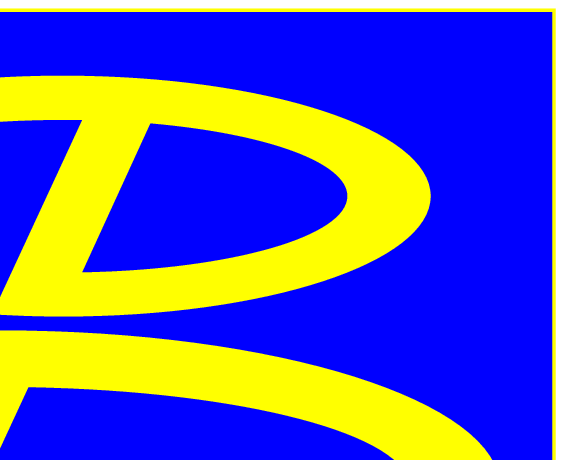}}

\vspace*{-3cm}
\begin{flushright}
BELLE-CONF-0344\\
KEK Preprint 2003-47\\
Belle Preprint 2003-14
\end{flushright}
\vspace*{1cm}

\def\bz{{B^0}}
\def\bzb{{\overline{B}{}^0}}
\def\kl{K_L^0}
\def\dE{{\Delta E}}
\def\mb{{M_{\rm bc}}}
\def\Dt{\Delta t}
\def\Dz{\Delta z}
\def\fol{f_{\rm ol}}
\def\fsig{f_{\rm sig}}
\newcommand{\sinbb}{{\sin2\phi_1}}

\newcommand{\ra}{\rightarrow}
\newcommand{\myindent}{\hspace*{2cm}}  
\newcommand{\fCP}{f_{CP}}
\newcommand{\ftag}{f_{\rm tag}}
\newcommand{\zCP}{z_{CP}}
\newcommand{\tCP}{t_{CP}}
\newcommand{\ttag}{t_{\rm tag}}
\newcommand{\cala}{{\cal A}}
\newcommand{\cals}{{\cal S}}
\newcommand{\dm}{\Delta m_d}
\newcommand{\dmd}{\dm}
\def\taubz{\tau_\bz}
\def\ks{{K_S^0}}
\newcommand{\btosqq}{b \to s\overline{q}q}
\newcommand{\btosss}{b \to s\overline{s}s}
\newcommand*{\dwl}{\ensuremath{{\Delta w_l}}}
\newcommand*{\fq}{\ensuremath{q}}
\title{\boldmath Measurement of Time-Dependent {\boldmath $CP$}-Violating 
       \\Asymmetries in $\bz \to \phi\ks$, $K^+K^-\ks$ and $\eta'\ks$ Decays}

\date{\today}

\affiliation{Budker Institute of Nuclear Physics, Novosibirsk}
\affiliation{Chiba University, Chiba}
\affiliation{University of Cincinnati, Cincinnati, Ohio 45221}
\affiliation{University of Frankfurt, Frankfurt}
\affiliation{Gyeongsang National University, Chinju}
\affiliation{University of Hawaii, Honolulu, Hawaii 96822}
\affiliation{High Energy Accelerator Research Organization (KEK), Tsukuba}
\affiliation{Hiroshima Institute of Technology, Hiroshima}
\affiliation{Institute of High Energy Physics, Chinese Academy of Sciences, Beijing}
\affiliation{Institute of High Energy Physics, Vienna}
\affiliation{Institute for Theoretical and Experimental Physics, Moscow}
\affiliation{J. Stefan Institute, Ljubljana}
\affiliation{Kanagawa University, Yokohama}
\affiliation{Korea University, Seoul}
\affiliation{Kyungpook National University, Taegu}
\affiliation{Institut de Physique des Hautes \'Energies, Universit\'e de Lausanne, Lausanne}
\affiliation{University of Ljubljana, Ljubljana}
\affiliation{University of Maribor, Maribor}
\affiliation{University of Melbourne, Victoria}
\affiliation{Nagoya University, Nagoya}
\affiliation{Nara Women's University, Nara}
\affiliation{National Kaohsiung Normal University, Kaohsiung}
\affiliation{National Lien-Ho Institute of Technology, Miao Li}
\affiliation{Department of Physics, National Taiwan University, Taipei}
\affiliation{H. Niewodniczanski Institute of Nuclear Physics, Krakow}
\affiliation{Nihon Dental College, Niigata}
\affiliation{Niigata University, Niigata}
\affiliation{Osaka City University, Osaka}
\affiliation{Osaka University, Osaka}
\affiliation{Panjab University, Chandigarh}
\affiliation{Peking University, Beijing}
\affiliation{Princeton University, Princeton, New Jersey 08545}
\affiliation{RIKEN BNL Research Center, Upton, New York 11973}
\affiliation{Saga University, Saga}
\affiliation{University of Science and Technology of China, Hefei}
\affiliation{Seoul National University, Seoul}
\affiliation{Sungkyunkwan University, Suwon}
\affiliation{University of Sydney, Sydney NSW}
\affiliation{Tata Institute of Fundamental Research, Bombay}
\affiliation{Toho University, Funabashi}
\affiliation{Tohoku Gakuin University, Tagajo}
\affiliation{Tohoku University, Sendai}
\affiliation{Department of Physics, University of Tokyo, Tokyo}
\affiliation{Tokyo Institute of Technology, Tokyo}
\affiliation{Tokyo Metropolitan University, Tokyo}
\affiliation{Tokyo University of Agriculture and Technology, Tokyo}
\affiliation{Toyama National College of Maritime Technology, Toyama}
\affiliation{University of Tsukuba, Tsukuba}
\affiliation{Utkal University, Bhubaneswer}
\affiliation{Virginia Polytechnic Institute and State University, Blacksburg, Virginia 24061}
\affiliation{Yokkaichi University, Yokkaichi}
\affiliation{Yonsei University, Seoul}
  \author{K.~Abe}\affiliation{High Energy Accelerator Research Organization (KEK), Tsukuba} 
  \author{K.~Abe}\affiliation{Tohoku Gakuin University, Tagajo} 
  \author{T.~Abe}\affiliation{High Energy Accelerator Research Organization (KEK), Tsukuba} 
  \author{I.~Adachi}\affiliation{High Energy Accelerator Research Organization (KEK), Tsukuba} 
  \author{H.~Aihara}\affiliation{Department of Physics, University of Tokyo, Tokyo} 
  \author{K.~Akai}\affiliation{High Energy Accelerator Research Organization (KEK), Tsukuba} 
  \author{M.~Akatsu}\affiliation{Nagoya University, Nagoya} 
  \author{Y.~Asano}\affiliation{University of Tsukuba, Tsukuba} 
  \author{T.~Aso}\affiliation{Toyama National College of Maritime Technology, Toyama} 
  \author{T.~Aushev}\affiliation{Institute for Theoretical and Experimental Physics, Moscow} 
% ...
\collaboration{The Belle Collaboration}

\begin{abstract}
  We present an improved measurement of $CP$-violation parameters
  in
  $\bz \to \phi\ks$, $K^+K^-\ks$, and $\eta'\ks$ decays
  based on a 140~fb$^{-1}$ data sample collected at the
  $\Upsilon(4S)$ resonance with
  the Belle detector at the KEKB energy-asymmetric $e^+e^-$ collider.
  One neutral $B$ meson is fully reconstructed in
  one of the specified decay channels,
  and the flavor of the accompanying $B$ meson is identified from
  its decay products.
  $CP$-violation parameters for each of the three
  modes are obtained from the asymmetries in the distributions of
  the proper-time intervals between the two $B$ decays.
  We find that the observed $CP$ asymmetry in the $B\to \phi \ks$ decay
  differs from the standard model (SM) expectation 
  by 3.5 standard deviations, while the other cases are consistent with
  the SM.
\end{abstract}


\maketitle

In the standard model (SM), $CP$ violation arises from an 
irreducible phase, the Kobayashi-Maskawa (KM) phase~\cite{bib:KM},
in the weak-interaction quark-mixing matrix. In
particular, the SM predicts $CP$ asymmetries in the time-dependent rates for $\bz$ and
$\bzb$ decays to a common $CP$ eigenstate $\fCP$~\cite{bib:sanda}. Recent measurements of the $CP$-violation
parameter $\sin2\phi_1$ by the Belle~\cite{bib:CP1_Belle}
and BaBar~\cite{bib:CP1_BaBar} collaborations established $CP$ violation
in $\bz \to J/\psi \ks$ and related decay modes~\cite{bib:CC},
which are governed by the $b \to c\overline{c}s$ transition,
at a level consistent with KM expectations.

Despite this success, many tests remain before one can conclude that the KM phase is
the only source of $CP$ violation.
The $\bz\to \phi\ks$ decay, which is dominated by the $\btosss$ transition,
is sensitive to new $CP$-violating phases from physics beyond the SM~\cite{bib:lucy}.
The other charmless decays $\bz\to K^+K^-\ks$ and $\bz\to \eta'\ks$,
which are mediated by $\btosss$, $s\overline{u}u$ and 
$s\overline{d}d$ transitions,
also provide additional information.
Since the SM predicts that measurements of $CP$ violation in these charmless modes 
should also yield $\sin 2\phi_1$ to 
a good approximation~\cite{bib:tree-penguin,bib:Garmash},
a significant deviation in the time-dependent $CP$ asymmetry in these 
modes from what is observed 
in $b \to c\overline{c}s$ decays would be evidence for a new $CP$-violating phase.

In the decay chain $\Upsilon(4S)\to \bz\bzb \to f_{CP}f_{\rm tag}$,
where one of the $B$ mesons decays at time $t_{CP}$ to a final state $f_{CP}$ 
and the other decays at time $t_{\rm tag}$ to a final state  
$f_{\rm tag}$ that distinguishes between $B^0$ and $\bzb$, 
the decay rate has a time dependence
given by~\cite{bib:sanda}
\begin{equation}
\label{eq:psig}
{\cal P}(\Delta{t}) = 
\frac{e^{-|\Delta{t}|/{\taubz}}}{4{\taubz}}
\biggl\{1 + \fq\cdot 
\Bigl[ \cals\sin(\dmd\Delta{t})
   + \cala\cos(\dmd\Delta{t})
\Bigr]
\biggr\},
\end{equation}
where $\taubz$ is the $B^0$ lifetime, $\dmd$ is the mass difference 
between the two $B^0$ mass
eigenstates, $\Delta{t}$ = $t_{CP}$ $-$ $t_{\rm tag}$, and
the $b$-flavor charge $\fq$ = +1 ($-1$) when the tagging $B$ meson
is a $B^0$ 
($\bzb$).
$\cals$ and $\cala$ are $CP$-violation parameters;
to a good approximation,
the SM predicts $\cals = -\xi_f\sin 2\phi_1$, where $\xi_f = +1 (-1)$ 
corresponds to  $CP$-even (-odd) final states, and $\cala =0$
for both $b \to c\overline{c}s$ and 
$b \to s\overline{s}s$ transitions. 
The present world-average $\cals$ value 
obtained from previous measurements by
the Belle~\cite{bib:Belle_sss} and BaBar~\cite{bib:BaBar_sss} collaborations
is within 2$\sigma$ of the SM expectation
for $\bz \to \eta'\ks$ and $K^+K^-\ks$, while
a 2.7$\sigma$ deviation exists for $\bz \to \phi\ks$.
Measurements with a larger data sample are required to resolve
this difference.

Belle's previous measurement
for $\bz \to \phi\ks$, $K^+K^-\ks$ and $\eta'\ks$
was based on a 78 fb$^{-1}$ data sample
containing 85 million $B\overline{B}$ pairs.
In this Letter, we report an improved measurement 
incorporating an additional 62 fb$^{-1}$ for a total of
140 fb$^{-1}$ (152 million $B\overline{B}$ pairs).
At the KEKB energy-asymmetric 
$e^+e^-$ (3.5 on 8.0~GeV) collider~\cite{bib:KEKB},
the $\Upsilon(4S)$ is produced
with a Lorentz boost of $\beta\gamma=0.425$ nearly along
the electron beamline ($z$).
Since the $B^0$ and $\bzb$ mesons are approximately at 
rest in the $\Upsilon(4S)$ center-of-mass system (cms),
$\Delta t$ can be determined from the displacement in $z$ 
between the $f_{CP}$ and $f_{\rm tag}$ decay vertices:
$\Delta t \simeq (z_{CP} - z_{\rm tag})/\beta\gamma c
 \equiv \Delta z/\beta\gamma c$.

The Belle detector~\cite{bib:Belle} is a large-solid-angle spectrometer
that includes a three-layer silicon vertex detector (SVD),
a 50-layer central drift chamber (CDC),
an array of aerogel threshold Cherenkov counters (ACC),
time-of-flight (TOF) scintillation counters,
and an electromagnetic calorimeter comprised of CsI(Tl) crystals (ECL)
located inside a superconducting solenoid coil
that provides a 1.5~T magnetic field.
An iron flux-return located outside of the coil is instrumented
to detect $\kl$ mesons and to identify muons (KLM).

We reconstruct $\bz$ decays to
$\phi\ks$ and $\eta'\ks$ final states for $\xi_f=-1$, and
$\bz\to K^+K^-\ks$ decays
that are a mixture of $\xi_f=+1$ and $-1$.
$K^+K^-$ pairs that are consistent with $\phi \to K^+K^-$ decay are excluded
from the $\bz \to K^+K^-\ks$ sample.
We find that the $K^+K^-\ks$ state is primarily $\xi_f=+1$;
a measurement of the $\xi_f=+1$ fraction with a 140~fb$^{-1}$ data set 
yields $1.03 \pm 0.15\mbox{(stat)}\pm0.05\mbox{(syst)}$, which
is consistent with the previous result~\cite{bib:Garmash}.
In the following determination of $\cals$ and $\cala$,
we fix $\xi_f=+1$ for this mode.
The intermediate meson states are reconstructed from the following decay chains:
$\eta'\to\rho^0 (\to \pi^+\pi^-) \gamma$ or $\eta'\to\pi^+\pi^-\eta (\to \gamma\gamma)$,
$\ks \to \pi^+\pi^-$, and $\phi\to K^+K^-$.
Candidate $\ks \to \pi^+\pi^-$ and $\phi \to K^+K^-$ decays are
selected with the same criteria as those used for the previous
branching fraction measurements~\cite{bib:Belle_phik}.
The $\ks$ selection is slightly changed from the previously published 
$CP$ asymmetry measurement~\cite{bib:Belle_sss} to improve the
$\ks$ purity.

Since the $\phi$ meson selection is effective in reducing background events,
we impose only minimal kaon-identification requirements.
We use more stringent kaon-identification requirements
to select non-resonant $K^+K^-$ candidates 
for the $\bz \to K^+K^-\ks$ decay~\cite{bib:Garmash}.
We reject $K^+K^-$ pairs that are consistent with 
$D^0 \to K^+K^-$, $\chi_{c0} \to K^+K^-$, or $J/\psi \to K^+K^-$ decay.
$D^+ \to \ks K^+$ candidates are also removed.
We use the same $\eta'$ selection criteria as those used in our 
previously published analyses~\cite{bib:Belle_sss,bib:Belle_etapks}. 

For reconstructed $B\to\fCP$ candidates, we identify $B$ meson decays using the
energy difference $\dE\equiv E_B^{\rm cms}-E_{\rm beam}^{\rm cms}$ and
the beam-energy constrained mass $\mb\equiv\sqrt{(E_{\rm beam}^{\rm cms})^2-
(p_B^{\rm cms})^2}$, where $E_{\rm beam}^{\rm cms}$ is
the beam energy in the cms, and
$E_B^{\rm cms}$ and $p_B^{\rm cms}$ are the cms energy and momentum of the 
reconstructed $B$ candidate, respectively.
The $B$ meson signal region is defined as 
$|\dE|<0.06$ GeV for $\bz \to \phi \ks$,
$|\dE|<0.04$ GeV for $\bz \to K^+K^-\ks$,
$|\dE|<0.06$ GeV for $\bz \to \eta'(\to \rho\gamma) \ks$, or
$-0.10$ GeV $< \dE <0.08$ GeV for $\bz \to \eta'(\to \pi^+\pi^-\eta) \ks$,
and $5.27~{\rm GeV}/c^2 <\mb<5.29~{\rm GeV}/c^2$ for all decays.
In order to suppress background from the $e^+e^- \rightarrow 
u\overline{u},~d\overline{d},~s\overline{s}$, or $c\overline{c}$
continuum,
we form signal and background
likelihood functions, ${\cal L}_{\rm S}$ and ${\cal L}_{\rm BG}$, 
from a set of variables that characterize the event topology,
and impose thresholds on the likelihood ratio 
${\cal L}_{\rm S}/({\cal L}_{\rm S}+{\cal L}_{\rm BG})$~\cite{bib:Belle_phik}. 
The threshold value depends both on the decay mode and 
on the flavor-tagging quality.

The $b$-flavor of the accompanying $B$ meson is identified
from inclusive properties of particles
that are not associated with the reconstructed $\bz \to \fCP$ 
decay~\cite{bib:CP1_Belle}.
We use two parameters, $\fq$ and $r$, to represent the tagging information.
The first, $\fq$, is already defined in Eq.~(\ref{eq:psig}).
The parameter $r$ is an event-by-event,
MC-determined flavor-tagging dilution factor
that ranges from $r=0$ for no flavor
discrimination to $r=1$ for unambiguous flavor assignment.
It is used only to sort data into six $r$ intervals.
The wrong tag fractions for the six $r$ intervals, 
$w_l~(l=1,6)$, and differences 
between $\bz$ and $\bzb$ decays, $\dwl$,
are determined from the data;
we use the same values
that were used for the $\sin 2\phi_1$ measurement~\cite{bib:BELLE-CONF-0353}.

The decay vertices of $\bz$ mesons are reconstructed using
tracks that have enough SVD hits.
The vertex position for the $\fCP$ decay is reconstructed
using charged tracks excluding pions from the $\ks$ decays.
The $\ftag$ vertex determination remains unchanged from the
previous publication~\cite{bib:Belle_sss},
and is described in detail elsewhere~\cite{bib:resol_nim}.

\begin{figure}
\resizebox{0.6\textwidth}{!}{\includegraphics{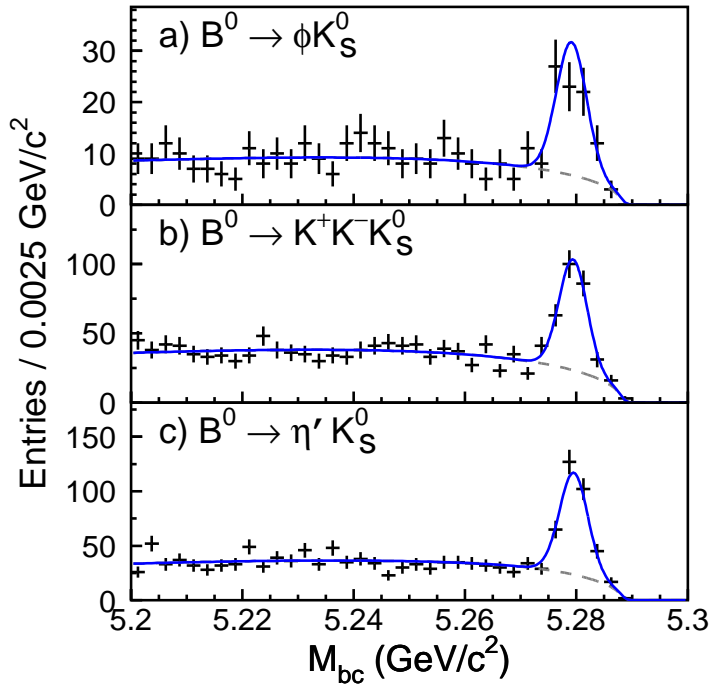}}
\caption{The $\mb$ distributions for
(a) $\bz\to\phi\ks$, (b) $\bz\to K^+K^-\ks$, and (c) $\bz\to\eta'\ks$
within the $\dE$ signal region.
Solid curves show the fit to signal plus background distributions,
and dashed curves show the background contributions.
The fit functions are the same as those used in the previous
publication~\cite{bib:Belle_sss}.
}\label{fig:mb}
\end{figure}

After flavor tagging and vertex reconstruction, we obtain the numbers of 
$\bz\to\fCP$ candidates, $N_{\rm ev}$,
listed in Table~\ref{tab:num}.
\begin{table}
\caption{
The numbers of reconstructed $\bz\to\fCP$ candidates
used for $\cals$ and $\cala$ determination, $N_{\rm ev}$, and the estimated
signal purity in the $\dE$-$\mb$ signal region for each $\fCP$ mode.}
\label{tab:num}
\begin{ruledtabular}
\begin{tabular}{llrl}
\multicolumn{1}{c}{Mode} & \multicolumn{1}{c}{$\xi_f$} 
                 & $N_{\rm ev}$ & \multicolumn{1}{c}{Purity} \\
\hline
$\phi\ks$   & $-1$        & 106 & $0.64\pm 0.10$\\
$K^+K^-\ks$ & $+1(100\%)$ & 361 & $0.55\pm 0.05$\\
$\eta'\ks$  & $-1$        & 421 & $0.58\pm 0.05$\\
\end{tabular}
\end{ruledtabular}
\end{table}
Figure~\ref{fig:mb} shows the $\mb$ distributions for the reconstructed $B$ candidates that have $\dE$
values within the signal region.    
We use events outside the signal region 
as well as a large MC sample to study the background components.
The dominant background comes from continuum events.
In addition, according to MC simulation, there is  a small ($\sim 3\%$) 
contamination from $B\overline{B}$ background events
in $\bz\to\eta'\ks~(\eta'\to\rho^0\gamma)$. 
The contributions from $B\overline{B}$ events are smaller for other 
decay modes.
The contamination of $K^+ K^- \ks$ events in the $\phi\ks$ sample 
($7.2\pm1.7$\%) is also small.
Finally, backgrounds from the $B^0 \rightarrow f_0(980) \ks$ decay, 
which has the opposite 
$CP$ eigenvalue to $\phi\ks$, are found to be small
($1.6^{+1.9}_{-1.5}$\%).  
The influence of these backgrounds
is treated as a source of systematic uncertainty.

We determine $\cals$ and $\cala$ for each mode by performing an unbinned
maximum-likelihood fit to the observed $\Dt$ distribution.
The probability density function (PDF) expected for the signal
distribution, ${\cal P}_{\rm sig}(\Dt;\cals,\cala,\fq,w_l,\dwl)$, 
is given by Eq.~(\ref{eq:psig}) incorporating
the effect of incorrect flavor assignment. The distribution is
convolved with the
proper-time interval resolution function $R_{\rm sig}(\Dt)$~\cite{bib:BELLE-CONF-0353},
which takes into account the finite vertex resolution. 
We determine the following likelihood value for each
event:
\begin{eqnarray}
P_i
&=& (1-\fol)\int_{-\infty}^{\infty}\biggl[
\fsig{\cal P}_{\rm sig}(\Dt')R_{\rm sig}(\Dt_i-\Dt') \nonumber \\
&+&(1-\fsig){\cal P}_{\rm bkg}(\Dt')R_{\rm bkg}(\Dt_i-\Dt')\biggr]
d(\Dt')  \nonumber \\
&+&\fol P_{\rm ol}(\Dt_i) 
\end{eqnarray}
where $P_{\rm ol}(\Dt)$ is a broad Gaussian function that represents
an outlier component with a small fraction $\fol$.
The signal probability $\fsig$ depends on the $r$ region and
is calculated on an event-by-event basis
as a function of $\dE$ and $\mb$~\cite{bib:Belle_sss}.
${\cal P}_{\rm bkg}(\Dt)$ is a PDF for background events,
which is modeled as a sum of exponential and prompt components, and
is convolved with a sum of two Gaussians $R_{\rm bkg}$.
All parameters in ${\cal P}_{\rm bkg} (\Dt)$
and $R_{\rm bkg}$ are determined by the fit to the $\Dt$ distribution of a 
background-enhanced control sample~\cite{bib:BBbg}; i.e. events away from the $\dE$-$\mb$ signal region.
We fix $\tau_\bz$ and $\dmd$ at
their world-average values~\cite{bib:PDG2003}.
The only free parameters in the final fit
are $\cals$ and $\cala$, which are determined by maximizing the
likelihood function
$L = \prod_iP_i(\Dt_i;\cals,\cala)$
where the product is over all events.
Table \ref{tab:result} gives 
the fit values of 
$-\xi_f\cals$ and $\cala$.
\begin{table}
\caption{Results of the fits to the $\Dt$ distributions.
The first errors are statistical and the second
errors are systematic.  The third error for the $K^+K^-\ks$ mode arises from
the uncertainty in the fraction of the $CP$-odd component.}
\label{tab:result}
\begin{ruledtabular}
\begin{tabular}{cll}
Mode &  \multicolumn{1}{c}{$-\xi_f\cals$ ($= \sinbb$ in the SM)} & 
\multicolumn{1}{c}{$\cala$ (= 0 in the SM)} \\
\hline
$\phi\ks$ & $-0.96\pm0.50^{+0.09}_{-0.11}$ & $-0.15\pm0.29\pm0.07$ \\
$K^+K^-\ks$ & $+0.51\pm0.26\pm0.05^{+0.18}_{-0.00}$ & $-0.17\pm0.16\pm0.04$ \\
$\eta'\ks$  & $+0.43\pm0.27\pm0.05$ & $-0.01\pm0.16\pm0.04$ \\
\end{tabular}
\end{ruledtabular}
\end{table}
These results are consistent with the previous results~\cite{bib:Belle_sss}
and supersede them.
We obtain values consistent with
the present world average of $\sin 2\phi_1 = +0.731\pm 0.056$~\cite{bib:PDG2003}
in the $\bz \to K^+K^-\ks$ and $\eta'\ks$ decays, while a negative value
is observed in $\bz \to \phi \ks$ decay.

We define the raw asymmetry in each $\Dt$ bin by
$A\equiv(N_{q\xi_f=-1}-N_{q\xi_f=+1})/(N_{q\xi_f=-1}+N_{q\xi_f=+1})$,
where $N_{q\xi_f=+1(-1)}$ is the number of observed candidates with $q\xi_f=+1(-1)$.
\begin{figure*}
\resizebox{!}{0.32\textwidth}{\includegraphics{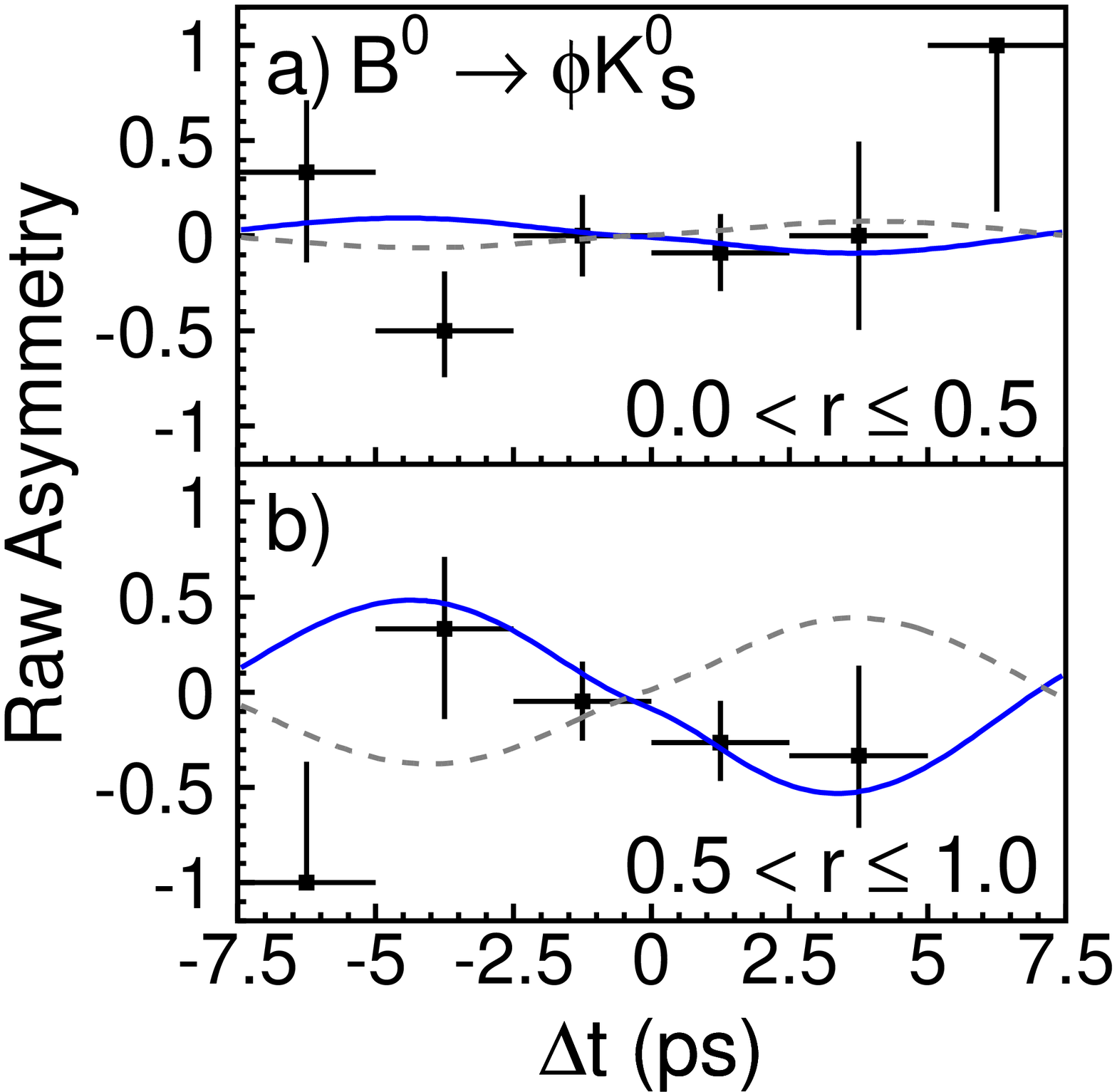}} 
\resizebox{!}{0.32\textwidth}{\includegraphics{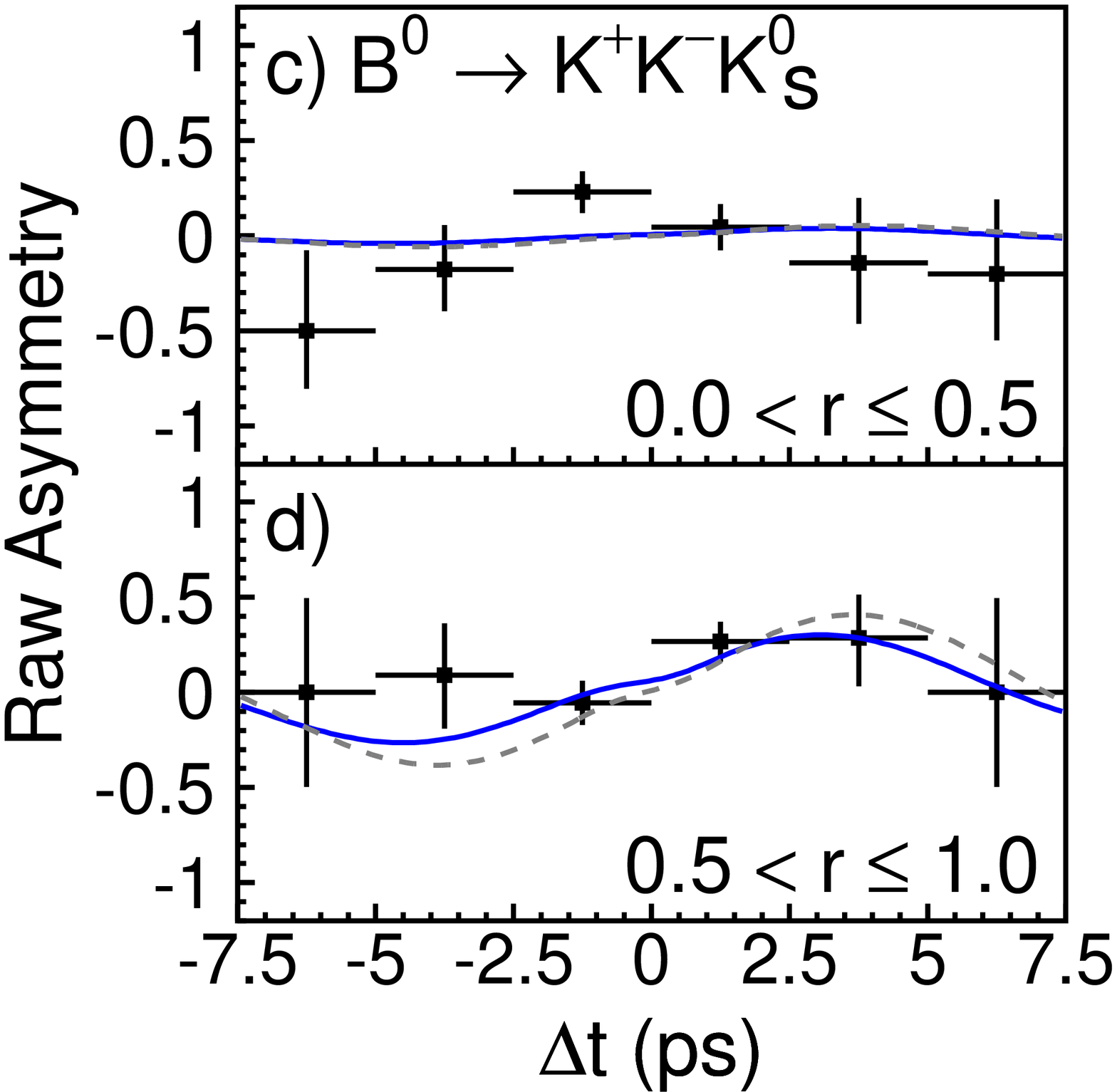}} 
\resizebox{!}{0.32\textwidth}{\includegraphics{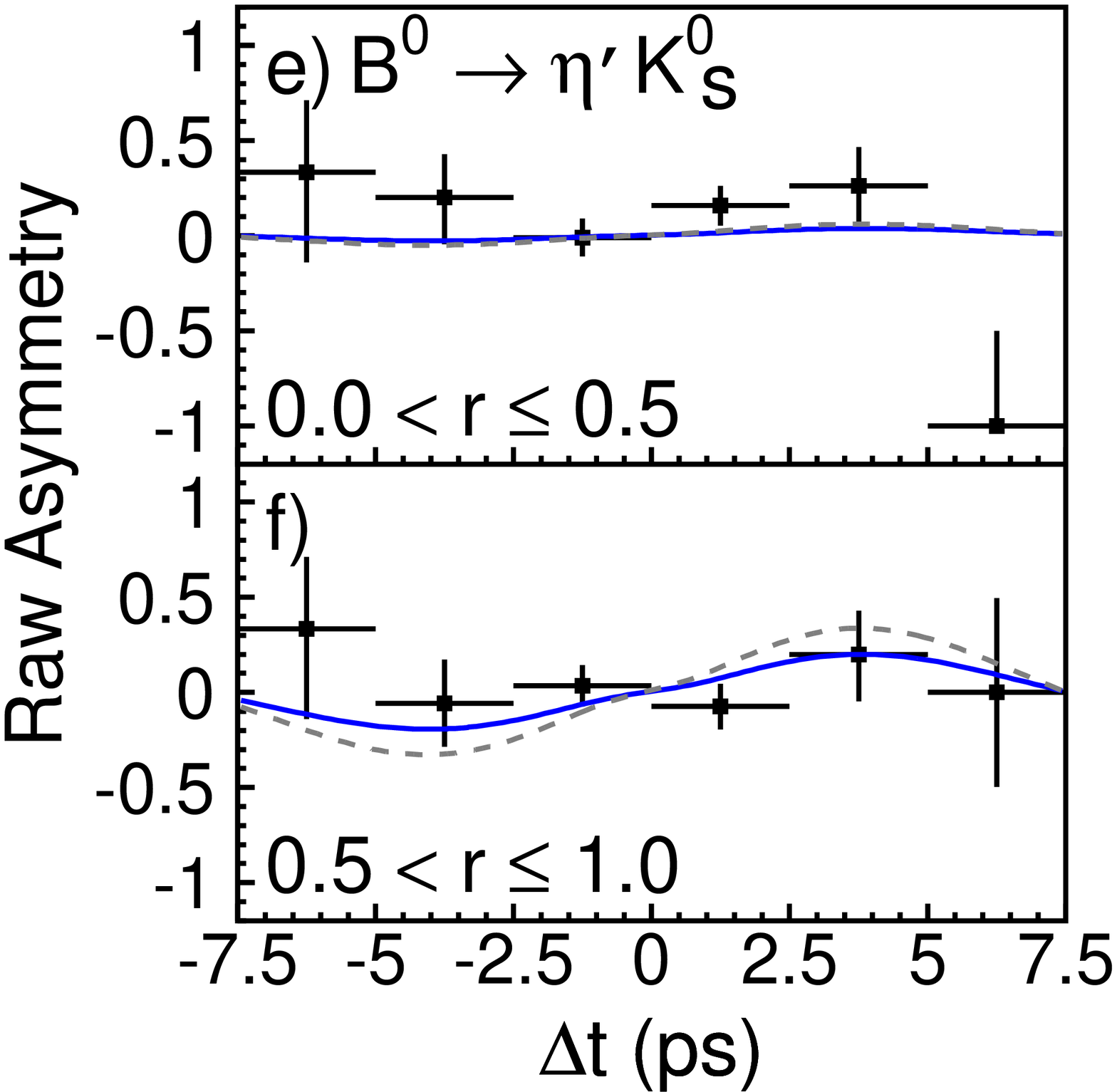}} 
\caption{
(a) The asymmetry, $A$, in each $\Dt$ bin for
$\bz\to\phi\ks$ with $0 < r \le 0.5$, 
(b) with $0.5 < r \le 1.0$,
(c) for $\bz\to K^+K^-\ks$ with $0 < r \le 0.5$, (d) with $0.5 < r \le 1.0$,
(e) for $\bz\to \eta'\ks$ with $0 < r \le 0.5$, and (f) with $0.5 < r \le 1.0$,
respectively. The solid curves show the result of the 
unbinned maximum-likelihood fit.
The dashed curves show the SM expectation with $\sin2\phi_1$ = +0.731 
and $\cala$ = 0.}
\label{fig:asym}
\end{figure*}
Figures~\ref{fig:asym}(a-f) show the raw asymmetries in two regions of the flavor-tagging
parameter $r$. While the numbers of events in the two regions are similar,
the effective tagging efficiency is much larger 
and the background dilution is smaller in the region $0.5 < r \le 1.0$.
The observed $CP$ asymmetry for $\bz \to \phi \ks$
in the region $0.5 < r \le 1.0$ [Fig.~\ref{fig:asym}(b)]
indicates the difference from the SM expectation (dashed curve). 
Note that these projections onto the $\Delta t$ axis do not take into
account event-by-event information (such as the signal fraction, the
wrong tag fraction and the vertex resolution), which is used in the
unbinned maximum-likelihood fit.

Fits to the same samples with 
$\cala$ fixed at zero yield
$-\xi_f\cals = -0.99\pm0.50$(stat) for $\bz\to\phi\ks$,
$-\xi_f\cals = +0.54\pm0.24$(stat) for $\bz\to  K^+K^-\ks$, and
$-\xi_f\cals =  +0.43\pm0.27$(stat) for $\bz\to\eta' \ks$.
Applying the same fit procedure,
we also obtain 
$\cals = -0.09\pm0.26$(stat), $\cala = +0.18 \pm 0.20$(stat) 
for $B^+ \to \phi K^+$ decay and
$\cals = +0.10\pm0.14$(stat), $\cala = -0.04 \pm 0.09$(stat) 
for $B^+ \to \eta' K^+$ decay.
Both results for the $\cals$ term are consistent with no $CP$ asymmetry, as expected. 

The dominant sources of systematic error for the $\bz \to \phi\ks$ mode are 
a possible fit bias for the input $\cals$ value near the physical boundary
($^{+0.06}_{-0.00}$ for $\cals$),
the uncertainties in
the $\bz \to K^+K^-\ks$ and $f_0(980) \ks$ background fractions
($^{+0.00}_{-0.08}$ for $\cals$ and $\pm 0.04$ for $\cala$),
in the other background fractions
($\pm0.05$ for $\cals$ and $\pm 0.04$ for $\cala$),
and in the vertex reconstruction
($\pm0.02$ for $\cals$ and $\pm0.05$ for $\cala$).
Other contributions come from uncertainties in
the background $\Delta t$ distribution,
wrong tag fractions, $\taubz$, and $\dmd$.
We add each contribution in quadrature to obtain the total systematic
uncertainty.
Systematic uncertainties from these sources
are also examined for the other modes. 
We find that the dominant sources are uncertainties
from the background fractions and from the vertex reconstruction.
%

We use the Feldman-Cousins 
frequentist approach~\cite{bib:FeldmanCousins}
to determine the statistical significance of the observed 
deviation from the SM expectation
in $\bz \to \phi \ks$. The procedure is described in detail
elsewhere~\cite{bib:Belle_phi2prd}.
From 1-dimensional confidence intervals for $\cals$
with $\cala$ set at zero, the case with $\cals = +0.731$ 
is ruled out at 99.95\% confidence level, equivalent
to 3.5$\sigma$ significance for Gaussian errors.
As a cross check, we varied the selection criteria for the $\bz \to \phi \ks$ decay
and repeated the analysis. We find no sizable change in the significance.

In summary, we have performed improved measurements of 
$CP$-violation parameters for $\bz \to \phi \ks$, $K^+K^-\ks$ and
$\eta'\ks$ decays.
These charmless decays
are sensitive to possible new $CP$-violating phases.
Our results for $\bz\to\eta'\ks$ and $K^+K^-\ks$ are consistent with those obtained for
$\bz \to J/\psi \ks$ and other decays governed by the $b \to c\overline{c}s$ transition. 
On the other hand, a $3.5\sigma$ deviation is observed for $\bz\to\phi\ks$.
The result suggests that there is a large $CP$-violating phase in
its decay amplitude, which cannot be explained by the SM.

\begin{acknowledgments}
We wish to thank the KEKB accelerator group for the excellent
operation of the KEKB accelerator.
We acknowledge support from the Ministry of Education,
Culture, Sports, Science, and Technology of Japan
and the Japan Society for the Promotion of Science;
the Australian Research Council
and the Australian Department of Education, Science and Training;
the National Science Foundation of China under contract No.~10175071;
the Department of Science and Technology of India;
the BK21 program of the Ministry of Education of Korea
and the CHEP SRC program of the Korea Science and Engineering Foundation;
the Polish State Committee for Scientific Research
under contract No.~2P03B 01324;
the Ministry of Science and Technology of the Russian Federation;
the Ministry of Education, Science and Sport of the Republic of Slovenia;
the National Science Council and the Ministry of Education of Taiwan;
and the U.S.\ Department of Energy.
\end{acknowledgments}


\begin{thebibliography}{999}
\bibitem{bib:CC}
Throughout this paper,
the inclusion of the charge conjugate decay mode is implied
unless otherwise stated.
\bibitem{bib:BBbg}
Parameters in ${\cal P}_{\rm bkg}(\Dt)$ and $R_{\rm bkg}$
for $B\overline{B}$ background events in $\bz\to\eta'\ks$ decay
are determined from MC simulation.
\end{thebibliography}
\end{document}